\documentclass[10pt,final,conference,letterpaper,romanappendices,twocolumn]{IEEEtran}
\IEEEoverridecommandlockouts
\usepackage{cite}
\usepackage{textcomp}
\usepackage{xcolor}
\usepackage{epstopdf}
\usepackage{epsfig}
\usepackage{amsthm}
\usepackage{amsmath,amssymb,amsfonts,amstext,amsbsy,amsopn,dsfont}
\usepackage{enumitem}
\usepackage{cases}
\usepackage{sublabel}
\usepackage{bigints}
\usepackage{sublabel}
\usepackage{array}
\usepackage{cuted}

\newtheorem{theorem}{\textbf{Theorem}}

\newcommand{\defn}{\triangleq}
\newcommand{\dif}{\mathrm{d}}

\begin{document}

\title{Toward Ubiquitous and Flexible Coverage of UAV-IRS-Assisted NOMA Networks\\
\thanks{The work of C.-H. Liu and M. A. Syed was supported in part by the U.S. National Science Foundation (NSF) under Award CNS-2006453 and in part by Mississippi State University under Grant ORED 253551-060702. The work of L. Wei was supported in part by the NSF under Award CNS-2006612.}
}

\author{\IEEEauthorblockN{Chun-Hung Liu and Md Asif Syed}
\IEEEauthorblockA{Department of Electrical \& Computer Engineering \\
Mississippi State University, MS, USA \\
e-mail: chliu@ece.msstate.edu}
\and
\IEEEauthorblockN{Lu Wei}
\IEEEauthorblockA{Department of Computer Science\\
Texas Tech University, Lubbock TX, USA \\
e-mail: luwei@ttu.edu}
}

\maketitle

\begin{abstract}
This paper studies how to achieve a high and flexible coverage performance of a large-scale cellular network that enables unmanned aerial vehicles (UAVs) for non-orthogonal multiple access (NOMA) transmission to simultaneously serve multiple users. The considered cellular network consists of a tier of base stations and a tier of UAVs. Each UAV is mounted with an intelligent reflecting surface (IRS) in order to serve as an aerial IRS reflecting signals between a base station and a user in the network. All the UAVs in the network are deployed based on a newly proposed three-dimensional (3D) point process that leads to a tractable and accurate analysis of the association statistics, which is traditionally difficult to analyze due to the mobility of UAVs. In light of this, we are able to analyze the downlink coverage of UAV-IRS-assisted NOMA transmission for two users and derive the corresponding coverage probabilities. Our coverage analyses shed light on the optimal allocations of transmit power between NOMA users and UAVs to accomplish the goal of ubiquitous and flexible NOMA transmission. We also conduct numerical simulations to validate our coverage analytical results while demonstrating the improved coverage performance achieved by aerial IRSs. 
\end{abstract}


\section{Introduction}
In the past decade, the performance and efficiency of wireless transmission have been significantly improved thanks to the technological advances enhancing the quality of received signals of communication systems. The multi-antenna technology, for example, is able to considerably boost the signal strength of a point-to-point wireless link by exploiting the spatial and temporal diversities. The cognitive radio technology can improve the efficiency of spectrum utilization. Moreover, the millimeter wave technology leads to a gigabit-level transmission rate in 5G systems. These technologies were developed by following the core idea of strengthening the desired signals on both the transmitter and receiver sides. As these technologies are becoming mature, there may not be much room for further significant improvement in wireless transmission. We are thus imperative to seek other non-traditional and feasible solutions to benefiting wireless transmission when embarking on the journey of investigating next-generation wireless networks. 

There is a frequently neglected approach that is able to directly improve wireless signal transmission. The idea is to mitigate the impact of various environmental factors, such as, blockage, topography, and weather on the transmission performance of wireless channels. Recently, such an approach has been more viable in light of the advance in manufacturing intelligent reflecting surface (IRSs). An IRS is a passive device that can be manipulated to alter and reflect its incident signals with a negligible power loss. Therefore, the deployment of IRSs in wireless networks enables a more controllable wireless transmission environment, thereby nullifying the uncontrollable nature of wireless channels~\cite{MAEHZLS20,YLXLXM21,EBIY2021}. Accordingly, how to efficiently and economically deploy IRSs to achieve a satisfactory coverage of all users in the wireless network becomes a challenging task. This is especially true when some special transmission techniques, such as NOMA, are employed in wireless networks. To investigate the impact of IRS deployment on the coverage of a wireless NOMA network, in this paper we propose a three-dimensional (3D) large-scale wireless network enabling UAVs mounted with an IRS to reflect NOMA signals from base stations to users on the ground. To the best of our knowledge, the coverage problem of large-scale wireless networks using UAV-IRS-assisted NOMA has not been addressed in the literature. 

There are a few recent works focusing on the special case of a single-cell wireless network using UAVs with IRSs as intelligent and energy-efficient mobile devices~\cite{MAMSSPAIDOYH21,STTHHE21,AJMADAAEIYAMS21,GKWCLZNDWKWK21}. There, a UAV is utilized to improve the channel quality by adjusting its hovering position between a base station and a user. In particular, reference \cite{MAMSSPAIDOYH21} studied downlink UAV-IRS-assisted communication, where the coverage probability and ergodic capacity were studied using an elevation-angle-dependent path-loss model and the bounds of the average signal-to-noise power ratio were also derived. In~\cite{STTHHE21}, an integrated downlink model of a UAV-IRS-assisted cellular network when considering three different modes (i.e., UAV-only, IRS-only, and integrated UAV-IRS modes) was proposed, where the corresponding outage probability, ergodic capacity, and energy efficiency were analyzed. Reference~\cite{AJMADAAEIYAMS21} investigated the symbol error rate and outage probability of multi-layer IRS-assisted UAV communication without perfect channel state information, and it showed that the communication performance between UAVs can be improved when using a large number of IRS elements for phase error mitigation. Although the aforementioned works pave the way to the understanding of UAV-assisted communication, their results may not be applicable to a large-scale multi-cell wireless network.  

The main contributions of this paper are summarized as follows. We first propose a large-scale 3D model of UAV-IRS-assisted NOMA network consisting of base stations (BSs) on the ground and UAVs in the sky. We then analyze the resulting statistics of the 3D UAV association scheme considering channel blocking indicating how densely UAVs and users should be deployed and distributed for a highly feasible UAV-IRS-assisted NOMA transmission. The maximum achievable coverage probability of a near user and a far user in the cell of a BS is derived in the scenario when the BS schedules a UAV to reflect signals to the far user. The derived coverage probability helps to design optimal power allocation policies and formulate an optimization problem of UAV positioning for a highly ubiquitous and flexible coverage of users in the network. More importantly, our analytical findings reveal the fact that UAV-IRS-assisted communication brings more flexibility for BSs when pairing NOMA users.

\section{System Model}
We consider a cellular network with a tier of BSs deployed on the ground, a tier of UAVs deployed in the sky, and users scattered on the ground. All the users are distributed by following a 2D homogeneous Poisson point process (HPPP) of density $\mu$ and all the BSs form an independent HPPP $\Phi_B$ of density $\lambda_B$, which can be written as the following set
\begin{align}
\Phi_B\defn\{B_i\in\mathbb{R}^2: i\in\mathbb{N}_+\},
\end{align}
where $B_i$ denotes BS $i$ and its ground location. Each user associates with its nearest BS. Without loss of generality, consider a typical BS $B_o$ located at the origin, where the following formulation and analyses will be expressed via the location of $B_o$. All the UAVs are deployed based on the following 3D point processes proposed in our previous work~\cite{CHLDCL21}:
\begin{align}
\Phi_D\defn\bigg\{ & D_j\in\mathbb{R}^2\times \left(0,\frac{\pi}{2}\right): D_j=(X_j,\Theta_j), X_j\in\mathbb{R}^2, \nonumber \\
&\Theta_j\in\left(0,\frac{\pi}{2}\right), j\in\mathbb{N}_+ \bigg\},
\end{align}
where $D_j$ denotes UAV $j$ and its 3D location, $X_j$ is the projection of $D_j$ on the ground, and $\Theta_j$ is the elevation angle from $B_o$ to $D_j$. All the projections of the UAVs (i.e., set $\{X_j\in\mathbb{R}^2: j\in\mathbb{N}_+\}$) are assumed to form an independent 2D HPPP of density $\lambda_D$. 

\begin{figure}[t!]
	\centering
	\includegraphics[height=2.25in,width=0.95\linewidth]{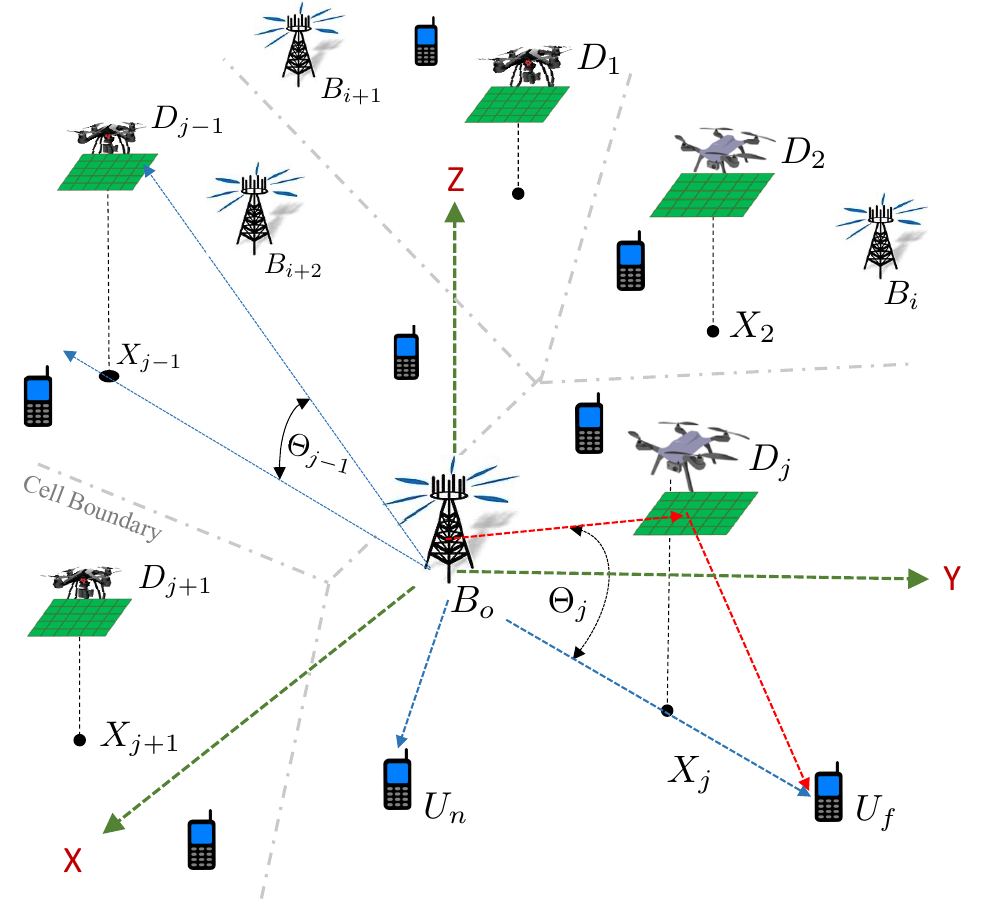}
	\caption{The proposed network model consists of a tier of BSs, a tier of UAVs mounted with an IRS, and users. A typical BS $B_o$ is located at the origin and it adopts NOMA transmission to serve a near user $U_n$ and a far user $U_f$. A UAV $D_j$ associating with $B_o$ is scheduled to reflect signals between $B_o$ and $U_f$. To minimize the path loss of the reflected signals between $B_0$ and $U_f$, $D_j$ hovers right above the middle point of the distance between $B_o$ and $U_f$ with an elevation angle $\Theta_j$, i.e., $\|X_j\|=\|D_j\|\cos(\Theta_j)=\frac{1}{2}\|U_f\|$.}
	\label{Fig:SystemModel}
	\vspace{-10pt}
\end{figure}

\subsection{UAV Channel Model and UAV Association}
A channel (link) between two spatial points is called a line-of-sight (LoS) channel if it is not visually blocked from one point to the other. Here, a low-altitude-platform (LAP) communication scenario is considered so that the LoS model of a 3D channel in~\cite{AHKSSL14} is adopted\footnote{In this paper, we assume that the 3D ground-to-air channels and the 3D air-to-ground channels are reciprocal and thereby they adopt the same LAP model.}, which gives rise to the LoS probability of a 3D link between UAV $j$ and any ground point $\mathsf{Y}_k\in\mathbb{R}^2$ given by
\begin{align}\label{Eqn:ProbLoS}
\rho(\Theta_{kj}) = \frac{1}{1+c_2\exp(-c_1\Theta_{kj})},
\end{align}
where $\Theta_{kj}\in(0,\frac{\pi}{2})$ is the elevation angle between $\mathsf{Y}_k$, and $D_j$, $c_1$ and $c_2$ are environment-dependent positive constants (for urban, rural, etc.). Each UAV associates with a BS in $\Phi_B$ that provides it with the strongest average power. As such, $D_j$ associates with BS $B_o$ whenever $B_o$ is satisfied with the following association scheme:
\begin{align}\label{Eqn:DefAssoUAV}
B_o = \arg\max_{i:B_i\in\Phi_B} \mathbb{E}\left[\frac{PH_{ij}L_{ij}}{\|B_i-D_j\|^{\alpha}}\bigg|B_i\right],
\end{align}
where $P$ is the transmit power of a BS, $\|B_i-D_j\|$ denotes the Euclidean distance between $B_i$ and $D_j$, $L_{ij}\in\{1,\eta\}$ is a Bernoulli random variable that is equal to $1$ if the link between $B_i$ and $D_j$ is a non-LoS (NLoS) one and $\eta$ otherwise, $\alpha>2$ is the path-loss exponent of a link, and $H_{ij}$ denotes the fading channel gain from $B_i$ to $D_j$. Throughout this paper, all the fading channel gains are assumed to be i.i.d. exponential random variables (RVs) with unit mean, e.g., $H_{ij}\sim\exp(1)$ for all $i,j\in\mathbb{N}_+$. Note that $\eta>1$ is used to model the enhanced channel gain of a 3D LoS link in contract to that of a 3D NLoS link so that it is referred to as the 3D LoS channel enhancement factor\footnote{To simplify the complexity of modeling and analysis, this paper assumes that all the 2D (ground-to-ground) channels are NLoS and adopt the same path-loss model as that used in the 3D NLoS channel model proposed in~\eqref{Eqn:DefAssoUAV}.}. Also note that $L_{ij}$ and $H_{ij}$ are independent since $L_{ij}$ is used to characterize whether or not the link between $D_j$ and $B_i$ is LoS  and it is completely determined for a given $B_i$ and $D_j$. Hence, the definition~\eqref{Eqn:DefAssoUAV} is reduced to
\begin{align}\label{Eqn:AssoUAV}
B_o &= \arg\max_{i:B_i\in\Phi_B} \frac{PL_{ij}\mathbb{E}[H_{ij}]}{\|B_i-D_j\|^{\alpha}}\nonumber\\
&=\arg \min_{i:B_i\in\Phi_B} L^{-\frac{1}{\alpha}}_{ij}\|B_i-D_j\|,
\end{align}
where the second equality is obtained because constants $P$ and $\mathbb{E}[H_{ij}]=1$ do not affect the association result. Note that the distribution of $L_{ij}$ depends on the elevation angle between $B_i$ and $D_j$ according to~\eqref{Eqn:ProbLoS} and all $L_{ij}$'s are assumed to be i.i.d. 

Using the expression of $B_o$ in~\eqref{Eqn:AssoUAV}, the numbers of users and UAVs become RVs, whose distributions are shown in the following theorem.
\begin{theorem}\label{Thm:StatisticsAsso}
Let $M$ and $N$ denote the number of the users and the number of the UAVs associating with BS $B_o$, respectively. The probability mass function (PMF) of $M$ can be accurately approximated by
\begin{align}
\hspace{-8pt} \mathbb{P}[M=m] &\approx \frac{\Gamma\left(m+\frac{7}{2}\right)}{m!\Gamma\left(\frac{7}{2}\right)}\left(\frac{2\mu}{7\lambda_B}\right)^m\left(1+\frac{2\mu}{7\lambda_B}\right)^{-(m+\frac{7}{2})}, \label{Eqn:PMFofUsers}
\end{align}
where $\Gamma(z)\defn \int_{0}^{\infty} x^{z-1}e^{-x}\dif x$ is the Gamma function. If the elevation angle $\Theta_{ij}$ between $B_i$ and $D_j$ is independent of the distance between $B_i$ and the projection $X_j$ of $D_j$ (i.e. $\|B_i-X_j\|$) and all $\Theta_{ij}$'s are i.i.d. for all $i,j\in\mathbb{N}_+$, the PMF of $N$ can be accurately approximated by
\begin{align}
\mathbb{P}[N=n] &\approx \frac{\Gamma\left(n+\frac{7}{2}\right)}{n!\Gamma\left(\frac{7}{2}\right)}\left(\frac{2\lambda_D}{7w_b\lambda_B}\right)^n\left(1+\frac{2\lambda_D}{7w_b\lambda_B}\right)^{-(n+\frac{7}{2})}, \label{Eqn:PMFofUAVs}
\end{align}
where $w_b$ is defined as
\begin{eqnarray}\label{Eqn:AssoWeight}
w_b&\defn& \mathbb{E}\left\{\cos^2(\Theta)\left[\rho(\Theta)\left(1-\eta^{\frac{2}{\alpha}}\right)+\eta^{\frac{2}{\alpha}}\right]\right\}\nonumber\\
&&\times \mathbb{E}\left\{\sec^{2}(\Theta)\left[\rho(\Theta)(1-\eta^{-\frac{2}{\alpha}})+\eta^{-\frac{2}{\alpha}}\right]\right\},
\end{eqnarray}
with the distribution of RV $\Theta$ being the same as that of $\Theta_{ij}$. 
\end{theorem}
\begin{IEEEproof}
See Appendix~\ref{App:ProofStatisticsAsso}.
\end{IEEEproof}
From Theorem~\ref{Thm:StatisticsAsso}, we see that $\mathbb{P}[M=0]\approx\mathbb{P}[N=0]\approx 0$ when $\mu$ and $\lambda_D$ are very large. Namely, each BS is almost surely associated with at least one UAV and one user if the densities of the users and the UAVs are sufficiently large. Thus, in this scenario we do not need to consider a cellular network having the phenomenon of \textit{void} cells in the network~\cite{CHLLCW16}\footnote{The cell of a BS is called a void cell if no users and UAVs associate with the BS.}. Moreover, each UAV is mounted with an IRS consisting of $R$ reflecting elements so that it essentially serves as an aerial IRS that helps reflect signals between a BS and users. For example, $D_j$ associating with $B_o$ uses an IRS to reflect the signals between $B_o$ and a user associating with $B_o$. An illustration of the network model is depicted in Fig.~\ref{Fig:SystemModel}. 

\subsection{Downlink UAV-IRS-Assisted NOMA Transmission}
Here, we investigate the downlink transmission performance from a BS to its users when a UAV associating with the BS is used to assist the downlink transmission via an IRS. Suppose the densities of the users and the UAVs are sufficiently large so that there are almost surely at least one UAV and two users associating with $B_o$ in the cell of $B_o$ according to the statistics of the 2D and 3D user association results in Theorem~\ref{Thm:StatisticsAsso}. We are interested in the scenario when $B_o$ is associated with multiple users and would like to use the \textit{power-domain} NOMA to simultaneously transmit different data streams to different users through the same frequency band. In each of the downlink transmission slots, $B_o$ schedules two users, i.e., a near user and a far user, in its cell for NOMA transmission, which also schedules UAV $D_j$ to reflect the signals from $B_o$ to the far user. Suppose the network is \textit{interference-limited} and each UAV is able to control its IRS to merely reflect the signals to its desired user so that the signals reflected do not interfere with any other users' signals in the network. 

Let $U_n$ and $U_f$ denote respectively the near user and the far user scheduled by $B_o$ in a transmission slot. Assuming each BS adopts different radio resource blocks to transmit different NOMA signals in its cell and each user is able to remove some NOMA interference via successive interference cancellation (SIC) of the received signals. Then, the \textit{achievable} signal-to-interference ratio (SIR) of the near user $U_n$ can be defined as\footnote{When $P_f>P_n$, $U_n$ is able to perform SIC so as to remove $I_{N,n}$ such that $\gamma_n$ increases. Otherwise, $I_{N,n}$ remains in~\eqref{Eqn:NearUserSIR} and $\gamma_n$ cannot thus be improved by SIC.}
\begin{align}\label{Eqn:NearUserSIR}
\gamma_n \defn   \frac{P_nG_{on}\|B_o-U_n\|^{-\alpha}}{\mathds{1}(P_f<P_n) I_{N,n}+I_{B,n}},
\end{align}
where $P_n$ is the power for the transmission of the desired signal of $U_n$, $G_{on}\sim\exp(1)$ denotes the fading channel gain of the link from $B_o$ to $U_n$, $I_{N,n} \defn P_fG_{on}\|B_o-U_n\|^{-\alpha}$ is the intra-cell NOMA interference received by $U_n$, $P_f$ is the power used to transmit the desired signal of $U_f$, $\mathds{1}(\mathcal{A})$ is the indicator function that equals one if event $\mathcal{A}$ is true and zero otherwise, $I_{B,n}\defn \sum_{i:B_i\in\Phi_B\setminus B_o}P G_{in} \|B_i-U_n\|^{-\alpha}$ denotes the inter-cell interference from the BSs in the network, and $G_{in}\sim\exp(1)$ is the fading channel gain from $B_i$ to $U_n$. We also notice that $P_f+P_n=P$.

Since UAV $D_j$ is scheduled by $B_o$ to reflect the signals from $B_o$ to $U_f$ and the fact that the path-loss model of a metasurface-based IRS is considered~\cite{MAEHZLS20}, the unit signal power reflected by $D_j$ and received by $U_f$ is expressed as
\begin{align}\label{Eqn:RecSigPowerFarUser}
W_f \defn H_fL_{oj}L_{jf}(\|B_o-D_j\|+\|D_j-U_f\|)^{-\alpha}.
\end{align} 
Here, $H_f$ is the \textit{equivalent} fading channel gain that characterizes channel fading of the reflected channel from $B_o$ to $U_f$ through $D_j$. Also, $H_f$ is assumed to be a Gamma RV with shape $R$ and rate $1$ (i.e., $H_f\sim\text{Gamma}(R,1)$) in that $D_j$ is able to control its IRS with $R$ reflecting elements such that all the $R$ signals reflected by its IRS can be coherently combined at $U_f$. As such, the achievable SIR of $U_f$ can be defined as
\begin{align}\label{Eqn:FarUserSIR}
\gamma_f \defn \frac{P_f[W_f+ G_{of}\|B_o-U_f\|^{-\alpha}]}{\mathds{1}(P_n<P_f)I_{N,f}+I_{B,f}},
\end{align}
where $G_{of}\sim\exp(1)$ is the fading channel gain from $B_o$ to $U_f$, $I_{N,f}\defn P_n[W_f+ G_{of}\|B_o-U_f\|^{-\alpha}]$ denotes the intra-cell NOMA interference received by $U_f$, $I_{B,f}\defn \sum_{i:B_i\in\Phi_B\setminus B_o} PG_{if}\|B_i-U_f\|^{-\alpha}$ is the inter-cell interference. Note that UAV $D_j$ is able to fly to the position whose projection is in the middle point of the straight path between $B_o$ and $U_f$. In this way, the distance between $B_o$ and $U_f$ through $D_j$ is minimal for a given elevation angle, as shown in Fig.~\ref{Fig:SystemModel}. As a result, the coverage probability of the far user can be further increased by the mobility of UAVs. In the following section, we will use $\gamma_n$ and $\gamma_f$ to first define the coverage probabilities of the near and far users before providing insights into power allocations between the two users and UAVs for an improved NOMA transmission. 

\section{Achievable Coverage of UAV-IRS-Assisted Downlink NOMA}
According to the definition~\eqref{Eqn:NearUserSIR}, for a minimum required SIR threshold of a successful decoding $\beta>0$, the coverage probabilities of the near user and the far user are defined as
\begin{align}\label{Eqn:DefCovProb}
c_n\defn\mathbb{P}\left[ \gamma_n\geq\beta \right]\quad\text{and}\quad c_f\defn\mathbb{P}\left[ \gamma_f\geq\beta \right],
\end{align}
respectively. The explicit expressions of $c_n$ and $c_f$ are given in the following theorem.
\begin{theorem}\label{Thm:CovProb}
The coverage probability $c_n$ of the near user is
\begin{align}\label{Eqn:NearCovProb}
c_n = \begin{cases}
\left[1+\Psi\left(\frac{2}{\alpha},\frac{P\beta}{(P_n-\beta P_f)}\right) \right]^{-1}, &P_n> \max\{1,\beta\}P_f\\
\left[1+\Psi\left(\frac{2}{\alpha},\frac{P\beta}{P_n}\right) \right]^{-1}, & P_n\leq P_f
\end{cases},
\end{align}
where the function $\Psi(x,y)$ for $x,y>0$ is defined as
\begin{align}
\Psi(x,y) \defn y^x \left(\frac{\pi x}{\sin(\pi x)}-\int_{0}^{y^{-x}}\frac{\dif z}{1+z^{\frac{1}{x}}}\right).
\end{align}
If the number of the reflecting elements of an IRS is sufficiently large (i.e., $R\gg 1$), the coverage probability of the far user defined in~\eqref{Eqn:DefCovProb} can be accurately approximated by~\eqref{Eqn:FarCovProb} as shown on top of next page, where $\tau_i$ is defined as

\begin{figure*}
	\begin{align}\label{Eqn:FarCovProb}
		c_f \approx \sum_{i=0}^{2} \mathbb{E}_{\Theta}\left\{\rho^i(\Theta)[1-\rho(\Theta)]^{2-i}\frac{\dif^{R-1}}{\dif t^{R-1}}\bigg(\frac{t^{R-1}}{(R-1)!}\prod_{k=1}^{2}\left[1+\frac{1}{k}\Psi\left(\frac{2}{\alpha},\frac{1}{t}\right)\right]^{-1}\bigg)\bigg|_{t=\tau_i}\right\},
	\end{align}
	\hrulefill
\end{figure*}

\begin{align}\label{Eqn:TauPara}
\tau_i =\begin{cases}
\frac{\eta^i(P_f-\beta P_n)}{\beta P}\cos^{\alpha}(\Theta), & P_f> \max\{1,\beta\}P_n\\
\frac{\eta^iP_f}{\beta P }\cos^{\alpha}(\Theta), & P_f \leq P_n
\end{cases},
\end{align}
with $\Theta$ having the same distribution as $\Theta_j$.
\end{theorem}
\begin{IEEEproof}
See Appendix~\ref{App:ProofCovProb}.
\end{IEEEproof}
A few interesting and crucial observations of Theorem~\ref{Thm:CovProb} can be drawn. First, $c_f$ can be greater than $c_n$ as long as the signals reflected from the UAV to the far user have been sufficiently enhanced by making $R$ large and positioning the UAV channels as an LoS one. Hence, both of the users have a chance to perform SIC for SIR improvement depending on the transmit power allocation\footnote{This is different from the regular NOMA transmission without assisting by an IRS because it needs to allocate more power to the signals of the far user and requests the near user to do SIC so that the SIRs of the near user and the far users are fairly close.}. Second, we can conclude that the optimal power allocation policy to maximize $c_n$ in~\eqref{Eqn:NearCovProb} is $P_n> (1+2\beta) P_f$, yet the optimal power allocation policy to maximize $c_f$ in~\eqref{Eqn:FarCovProb} is $P_f> (1+2\beta) P_n$. These two power control policies are opposite. Nonetheless, it is better to choose the policy that benefits $c_n$ (i.e., $P_n> (1+2\beta) P_f$) since $c_f$ can also be significantly improved by the UAV in addition to the optimal power allocation. Third, the result~\eqref{Eqn:FarCovProb} clearly shows how an aerial IRS improves $c_f$ and how the position of a UAV affects $c_f$. When there is no UAV reflecting the signals from $B_o$ to $U_f$, $c_f$ will decrease significantly because $R$, $\eta$, and $\Theta$ in~\eqref{Eqn:FarCovProb} are reduced to zero, one, and zero, respectively. Besides, $c_f$ will also decrease if a UAV is positioned at a large elevation angle such that $\rho(\Theta)$ is small. Finally, we should maintain $\eta^i\cos^{\alpha}(\Theta)>1$ in~\eqref{Eqn:TauPara} almost surely in order for the aerial RIS to enjoy the improved NOMA transmission. Namely, the elevation angle of a UAV should be bounded as $0<\Theta < \cos^{-1}(\eta^{-\frac{2}{\alpha}})$. Consequently, we are able to formulate the following optimization problem for a fixed $\Theta$ in~\eqref{Eqn:FarCovProb}
\begin{align}\label{Eqn:OptAngle}
\max_{\Theta} c_f  (\Theta)\,\,\,\text{s.t. } 0<\Theta < \cos^{-1}(\eta^{-\frac{2}{\alpha}})
\end{align}
to maximize $c_f$. This optimization problem is not convex due to the complicated form of $c_f$ in~\eqref{Eqn:FarCovProb}, and thereby we resort numerical approaches for the solutions. In the following section, we will provide some numerical results to validate these above findings. 

\section{Numerical Results and Discussions}

\begin{table}[!t] 
	\centering
	\caption{Network Parameters for Simulation~\cite{AHKSSL14}}\label{Tab:SimPara}
	\begin{tabular}{|c|c|}
		\hline Transmit Power of BSs  (W)  $P$  & $30$ \\ 
		\hline Density of $\Phi_B$ (BSs/m$^2$) $\lambda_B$  & $1\times 10^{-5}$   \\ 
		\hline Density of $\Phi_D$ (UAV/$m^2$) $\lambda_D$  & $1\times 10^{-4}$   \\ 
		\hline Number of Reflecting Elements $R$ & $8$, $16$, $32$ (see figures) \\ 
		\hline Path-loss Exponent $\alpha$ & $3$\\ 
		\hline Parameters $(c_1,c_2)$ in \eqref{Eqn:ProbLoS} for Suburban & $(24.5811,39.5971)$ \\
		\hline 3D LoS Channel Enhancement Factor $\eta$ & $2.5$ \\
		\hline SIR Threshold $\beta$  & $0.5$ \\ 
		\hline
	\end{tabular}
	\vspace{-6pt} 
\end{table}

\begin{figure}[t!]
	\vspace{-10pt}
	\centering
	\includegraphics[height=2.65in,width=1\linewidth]{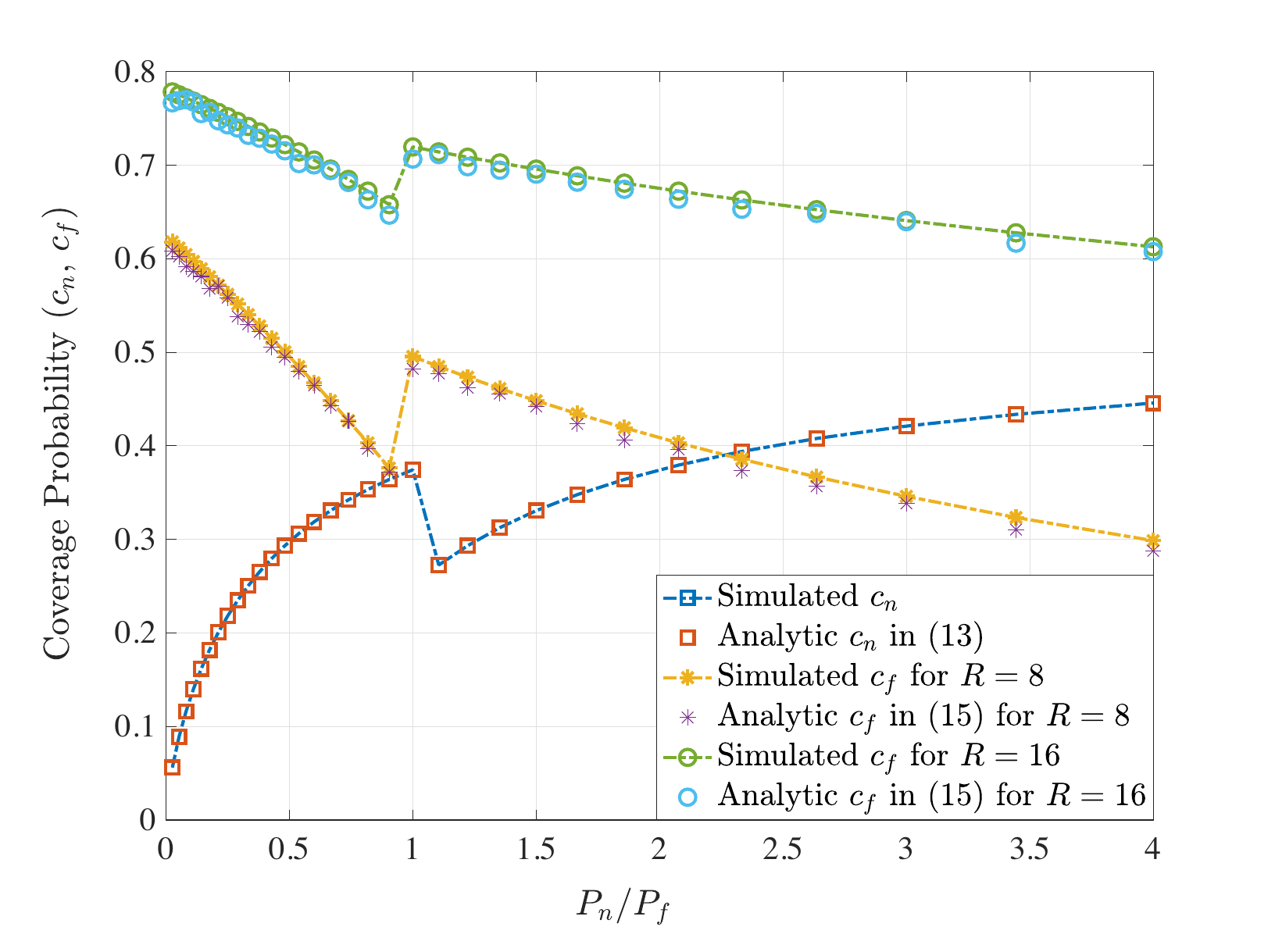}
	\caption{Simulation results of the coverage probabilities $c_n$ and $c_f$ for different numbers of the reflecting elements of an IRS. The elevation angle of the UAV is controlled at $\Theta=15^{\circ}$.}
	\label{Fig:CovProbPowRat}
	\vspace{-5pt}
\end{figure}

\begin{figure}[t!]
		\vspace{-10pt}
	\centering
	\includegraphics[height=2.65in,width=1\linewidth]{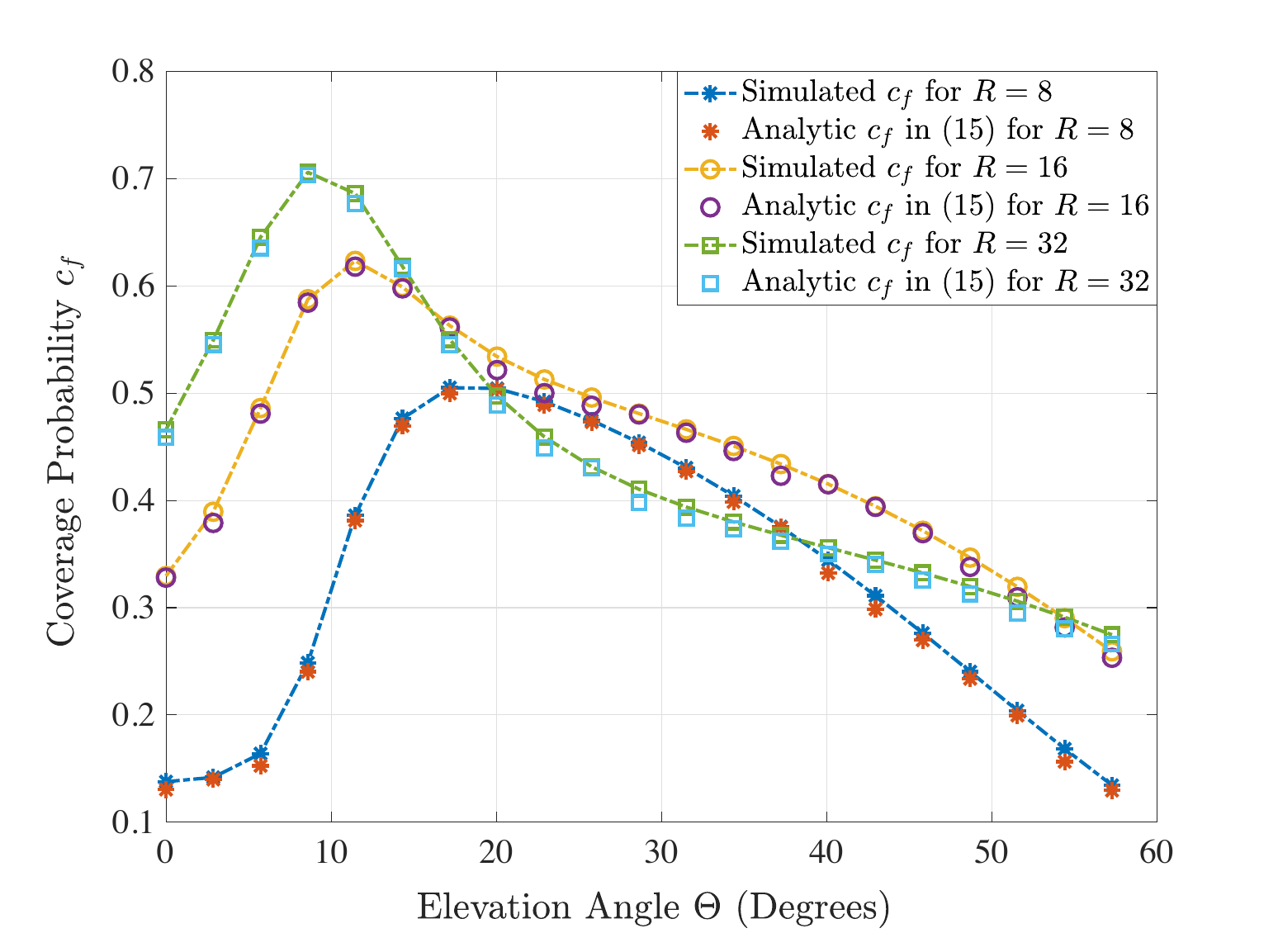}
	\caption{Simulation results of the coverage probabilities $c_n$ and $c_f$ when the UAV hovers at various elevation angle and the optimal power allocation policy $P_n\geq 2P_f$ is adopted.}
	\label{Fig:CovProbAngle}
	\vspace{-5pt}
\end{figure}

In this section, numerical simulation are provided to validate the coverage analyses of the previous section. To simplify the simulation scenarios, the elevation angle from a BS to a UAV is considered to be deterministic when the BS schedules the UAV to reflect its signals to a far user. The main network parameters used for simulation are listed in Table~\ref{Tab:SimPara}. We first present the numerical outcomes of the coverage probabilities $c_n$ and $c_f$ in Fig~\ref{Fig:CovProbPowRat} for two different numbers of $R$, where the elevation angle of the UAV is controlled at $\Theta=15^{\circ}$. As can be seen in the figure, all the analytic results match well with their corresponding simulated results, demonstrating the accuracy of the expressions~\eqref{Eqn:NearCovProb} and~\eqref{Eqn:FarCovProb}. The coverage probability $c_f$ significantly increases as $R$ increases from $8$ to $16$, as shown in the figure. In addition, in this simulation $c_f$ is much higher than $c_n$ due to the reflected transmission from the UAV. This observation reveals an important fact, that is, using UAVs with an RIS can remarkably boost the signal strength of users anywhere in the network so that the BS enjoys more flexibility of selecting users to maximize the transmission performance. Moreover, the optimal power allocation policy observed in the previous section, i.e., adopting  $P_n>(1+2\beta)P_f$ when $c_f>c_n$, can be observed in the figure. Since $\beta=0.5$, the optimal power allocation policy is $P_n/P_f>2$ and we can clearly see that the values of $c_n$ when $P_n/P_f>2$ are larger than those of $c_n$ when $P_n/P_f\leq 2$. Note that although using this policy may not always benefit $c_f$, we can always increase $c_f$ by adding more reflecting elements of an IRS on a UAV as well as controlling the UAV to improve the channels, as illustrated in Fig.~\ref{Fig:CovProbAngle}.

Fig. 3 shows how the coverage probability $c_f$ varies with the elevation angle of the UAV and the number of the reflecting elements of an IRS. As seen in the figure, $c_f$ improves as $R$ increases and there exists an optimal elevation angle that maximizes $c_f$ for each of the three curves. For example, the three optimal elevation angles for $R=8, 16, 32$ is about $9.2^{\circ}$, $12.2^{\circ}$, and $19.5^{\circ}$, which are all smaller than the upper bound $\cos^{-1}(\eta^{-2/\alpha})=57.12^{\circ}$ given in~\eqref{Eqn:OptAngle}. Hence, the optimization problem in~\eqref{Eqn:OptAngle} admits a feasible solution even though it is not convex, as pointed out in the previous section.  

\section{Conclusion}
This paper focuses on the study of how to effectively and significantly improve the coverage of users served by a BS with IRS-assisted NOMA transmission. The statistics of the 3D user association scheme for UAVs that considers LoS and NLoS channels has been analyzed. The results indicate how densely UAVs should be deployed and how the users should be distributed to ensure an almost surely feasible UAV-IRS-assisted NOMA transmission for a given density of BSs. Each BS is able to adopt NOMA to simultaneously serve a near user and a far user and to schedule a UAV to reflect signals to the far user. The maximum achievable coverage probabilities of the two users are also derived and analyzed. These results provide insights into optimal transmit power allocation between the two users as well as the optimal positioning of UAVs for NOMA transmission anywhere in the network. Our study also sheds light on the fact that UAV-IRS assisted communication brings BSs with more flexibility when pairing NOMA users in the cell of a BS.

\appendix
\numberwithin{equation}{section}
\setcounter{equation}{0}
\subsection{Proof of Theorem~\ref{Thm:StatisticsAsso}}\label{App:ProofStatisticsAsso}
Since each user associates with its nearest BS, the result in~\eqref{Eqn:PMFofUsers} can be readily obtained from Lemma 1 in~\cite{CHLKLF16}. To show the outcome in~\eqref{Eqn:PMFofUAVs}, we first need to rewrite~\eqref{Eqn:AssoUAV} as
\begin{align*}
B_o &= \arg \min_{i:B_i\in\Phi_B} L^{-\frac{2}{\alpha}}_{ij}\|B_i-D_j\|^2\\
&\stackrel{(a)}{=} \arg \max_{i:B_i\in\Phi_B} L_{ij}\sec^{-\alpha}(\Theta_{ij})\|B_i-X_j\|^{-\alpha}\\
&\stackrel{(b)}{=} \arg \max_{i:B_i\in\Phi_B} L_{ij}\cos^{\alpha}(\Theta_{ij})\|B_i\|^{-\alpha},
\end{align*}
where (a) is obtained due to $\|B_i-D_j\|=\sec(\Theta_{ij})\|B_i-X_j\|$ and (b) follows by considering $X_j$ as the origin and the fact that the statistic property evaluated at any point in an HPPP is the same according to the Slivnyak theorem~\cite{DSWKJM13}. If $\Theta_{ij}$ is independent of $\|B_i-X_j\|$, we can apply the result of Lemma 1 in~\cite{CHLKLF16} to find $w_b$ in~\eqref{Eqn:AssoWeight} as follows
\begin{align*}
w_b =&\mathbb{E}\left[\left(L_{ij}\cos^{\alpha}(\Theta_{ij})\right)^{\frac{2}{\alpha}}\right]\mathbb{E}\left[\left(L_{ij}\cos^{\alpha}(\Theta_{ij})\right)^{-\frac{2}{\alpha}}\right]\\
=& \mathbb{E}\left[L^{\frac{2}{\alpha}}\cos^2(\Theta)\right]\mathbb{E}\left[L^{-\frac{2}{\alpha}}\sec^{2}(\Theta)\right],
\end{align*}
which equals to~\eqref{Eqn:AssoWeight} by employing $\rho(\cdot)$ in~\eqref{Eqn:ProbLoS} of the above expression since all $L_{ij}\cos^{\alpha}(\Theta_{ij})$'s are independent.

\subsection{Proof of Theorem~\ref{Thm:CovProb}}\label{App:ProofCovProb}
According to~\eqref{Eqn:NearUserSIR} and~\eqref{Eqn:DefCovProb}, if $P_n>P_f$, the near user cannot perform SIC to remove $I_{N,n}$, and thereby $c_n$ are written as follows:
\begin{align}
	&c_n= \mathbb{P}\left[\gamma_n\geq\beta| P_n>P_f\right] =  \mathbb{P}\left[\frac{P_nG_n\|U_n\|^{-\alpha}}{ I_{N,n}+I_{B,n}}\geq\beta \right] \nonumber\\
	&=  \mathbb{P}\left[ I_{B,n}\leq \left(\frac{1}{\beta}-\frac{P_f}{P_n}\right)^+ P_nG_n\|U_n\|^{-\alpha}\right]\nonumber\\
	&= \mathbb{P}\left[\sum_{i:B_i\in\Phi_B\setminus B_o}\frac{P G_{in}\|U_n\|^{\alpha}}{P_n\|B_i-U_n\|^{\alpha}} \leq \left(\frac{1}{\beta}-\frac{P_f}{P_n}\right)^+ G_n\right]\nonumber\\
	&\stackrel{(a)}{=} \mathbb{E}\left[\exp\left\{ -\frac{P\beta}{(P_n-\beta P_f)^+} \sum_{i:B_i\in\Phi_B\setminus B_o}\frac{G_{in}\|U_n\|^{\alpha}}{\|B_i-U_n\|^{\alpha}} \right\}\right]\nonumber\\
	&\stackrel{(b)}{=} \left[1+\Psi\left(\frac{2}{\alpha},\frac{P\beta}{(P_n-\beta P_f)^+}\right) \right]^{-1}, \label{App:Eqn:CovProb}
\end{align}
where $x^+\defn\max\{0,x\}$, $(a)$ follows from the fact $G_n\sim\exp(1)$ and $(b)$ is obtained by using the result $\|U_n\|^2\sim\exp(\pi\lambda_B)$ before applying the derivation techniques devised in~\cite{CHLKLF16} based on the Probability Generating Functional (PGFL) of an HPPP. Hence, $c_n$ in~\eqref{Eqn:NearCovProb} for the case of $P_n>P_f$ is found. When $P_f\geq P_n$, the near user can perform SIC to remove $I_{N,n}$ from $\gamma_n$ so that the coverage probability of the near user in~\eqref{Eqn:NearCovProb} for the case of $P_f\geq P_n$ can be readily obtained by substituting $P_f=0$ into the coverage probability~\eqref{App:Eqn:CovProb}. 

To derive $c_f$, we first rewrite $W_f$ in~\eqref{Eqn:RecSigPowerFarUser} as $W_f = H_f L_{oj}L_{jf}\left(\sec(\Theta)\|U_f\|\right)^{-\alpha}$ by considering UAV $D_j$ hovering above the middle point of the straight path between $B_o$ and $U_f$ with an elevation angle $\Theta_j$. Then, assuming $R$ is large enough such that $W_f=H_f L_{oj}L_{jf}\left(\sec(\Theta_j)\|U_f\|\right)^{-\alpha}$ is almost surely much larger than $ G_f\|U_f\|^{-\alpha}$ (i.e., $W_f\gg G_f\|U_f\|^{-\alpha}$). Accordingly, we can accurately express $c_f$ in~\eqref{Eqn:DefCovProb} by neglecting $G_f\|U_f\|^{-\alpha}$ and conditioning on $L_{oj}L_{jf}$ for the case of $P_n<P_f$, $B_0=\mathbf{0}$ as follows:
\begin{align}
c_f \approx & \mathbb{P}\left[ \frac{P_fW_f}{P_nW_f+I_{B,f}} \geq \beta \right]= \mathbb{P}\left[ W_f \leq  \frac{\beta I_{B,f}}{\left(P_f-\beta P_n\right)^+} \right] \nonumber\\
=& \mathbb{P}\left[H_f\geq \frac{\beta I_{B,f}\sec^{\alpha}(\Theta_j)\|U_f\|^{\alpha}}{(P_f-\beta P_n)^+L_{oj}L_{jf}}\right]\nonumber\\
\stackrel{(c)}{=} & \frac{\dif^{R-1}}{\dif t^{R-1}}\bigg\{\frac{t^{R-1}}{(R-1)!}\mathbb{E}\bigg[e^{-\pi\lambda_B\Psi\left(\frac{2}{\alpha},\frac{1}{t}\right)\|U_f\|^2}\bigg]\bigg\}\bigg|_{t=t_o}, \label{Eqn:APPpf}
\end{align}
where $(c)$ is obtained by applying the derivation techniques in the proof of Proposition 1 in~\cite{CHLDCL18} and $t_o=\frac{(P_f-\beta P_n)^+L_{oj}L_{jf}}{\beta P\sec^{\alpha}(\Theta_j)}$. Furthermore, since $\|U_f\|\geq \|U_n\|$ and we know $\mathbb{E}[\exp(-\pi\lambda_Bs\|U_f\|^2)]=\frac{2}{(s+2)(s+1)}$ for $s>0$ according to the proof of Proposition 1 in~\cite{CHLDCL18}, we can infer the following result:
\begin{align*}
&\mathbb{E}\bigg[e^{-\pi\lambda_B\Psi\left(\frac{2}{\alpha},\frac{1}{t}\right)\|U_f\|^2}\bigg]=\prod_{k=1}^{2}\left[1+\frac{1}{k}\Psi\left(\frac{2}{\alpha},\frac{1}{t}\right)\right]^{-1}.
\end{align*}
Substituting the above result into~\eqref{Eqn:APPpf} yields
\begin{align*}
c_f\approx\frac{\dif^{R-1}}{\dif t^{R-1}}\bigg\{\frac{t^{R-1}}{(R-1)!}\prod_{k=1}^{2}\left[1+\frac{1}{k}\Psi\left(\frac{2}{\alpha},\frac{1}{t}\right)\right]^{-1}\bigg\}\bigg|_{t=t_0}.
\end{align*}
Finally, considering the RV $L_{oj}L_{jf}\in\{1,\eta, \eta^2\}$, where the independence between $L_{oj}$ and $L_{jf}$ leads to the expression of $c_f$ in~\eqref{Eqn:FarCovProb} for the case of $P_n>P_f$. For the case $P_f\leq P_n$, the substitution $P_n=0$ in~\eqref{Eqn:FarCovProb} for the case of $P_n>P_f$ gives rise to the expression of $c_f$ in~\eqref{Eqn:FarCovProb}.


\bibliographystyle{IEEEtran}
\bibliography{IEEEabrv,Ref_UAV-IRS-NOMA} 

\end{document}